\documentstyle[sprocl,epsfig]{article}

\def\PRD #1 #2 #3 {{\em Phys. Rev.} D {\bf#1},\ (#3) #2}
\def\PRL #1 #2 #3 {{\em Phys. Rev. Lett.} {\bf#1},\ (#3) #2}
\def\PLB #1 #2 #3 {{\em Phys. Lett.} B {\bf#1},\ (#3) #2}
\def\NPB #1 #2 #3 {{\em Nucl. Phys.} {\bf B#1},\ (#3) #2}
\def\ZPC #1 #2 #3 {{\em Z. Phys.} C {\bf#1},\ (#3) #2}
\def\EPJ #1 #2 #3 {{\em Euro. Phys. J.} C {\bf#1},\ (#3) #2}
\def\JHEP #1 #2 #3 {{\em JHEP} {\bf#1},\ (#3) #2}
\def\IJMP #1 #2 #3 {{\em Int. J. Mod. Phys.} A {\bf#1},\ (#3) #2}
\def\MPL #1 #2 #3 {{\em Mod. Phys. Lett.} A {\bf#1},\ (#3) #2}
\def\PTP #1 #2 #3 {{\em Prog. Theor. Phys.} {\bf#1},\ (#3) #2}
\def\PR #1 #2 #3 {{\em Phys. Rep.} {\bf#1},\ (#3) #2}
\def\RMP #1 #2 #3 {{\em Rev. Mod. Phys.} {\bf#1},\ (#3) #2}
\def\PRold #1 #2 #3 {{\em Phys. Rev.} {\bf#1},\ (#3) #2}
\def\IBID #1 #2 #3 {{\it ibid.} {\bf#1},\ (#3) #2}

\def\be{\begin{eqnarray}}
\def\ee{\end{eqnarray}}

\begin{document}

\title{Neutral Higgs Boson Production and CP Violation at LC\\}

\author{ Kang Young Lee \footnote{Conference Speaker} }

\address{
Department of Physics, Korea Advanced Institute of Science and Technology,
\\
Taejon 305-701, Korea}

\author{C.S. Kim, Chaehyun Yu }

\address{
Department of Physics and IPAP, Yonsei University, Seoul 120-749, Korea}

\maketitle\abstracts{
We study the neutral Higgs boson production
at the $e^- e^+$ linear collider
in the two Higgs doublet model with the CP violation.
The CP-even and CP-odd scalars are mixed in this model,
which affects the production process of neutral Higgs boson.
}

\section{Introduction}

Conventional wisdom is that the standard model (SM)
is not the final theory but just an effective theory
of a fundamental structure.
The two Higgs-doublet (2HD) model is one of the simplest
extension of the SM, which consist of two SU(2) scalar doublets.
A discrete symmetry is often introduced to avoid a dangerous
flavour-changing neutral currents (FCNC) in the 2HD model,
which forbids the CP violating terms in the Higgs potential.
If we abandon the discrete symmetry,
complex Higgs self-couplings exist in general, and consequently
the explicit and spontaneous CP violation is allowed in the Higgs sector.
In that case,
Higgs-mediated FCNC are present at tree level and
we must consider a way to suppress the FCNC, e.g.
to assume an approximate flavour symmetry.
Its phenomenological implicatiions have been widely studied
\cite{wu,hall,lee,huang,tt}.

In this work, we focus on the Higgs-gauge sector
in the general 2HD model with CP violation
and study the neutral Higgs boson production 
at the $e^- e^+$ Linear Colliders (LC).

\section{The Model}

The general Higgs potential of the 2HD model is given by
\be
V &=& \frac{1}{2} \lambda_1 (\phi_1^\dagger \phi_1)^2
   + \frac{1}{2} \lambda_2 (\phi_2^\dagger \phi_2)^2
   + \lambda_3 (\phi_1^\dagger \phi_1)(\phi_2^\dagger \phi_2)
   + \lambda_4 (\phi_1^\dagger \phi_2)(\phi_2^\dagger \phi_1)
\nonumber
\\
&& + \frac{1}{2} [ \lambda_5 (\phi_1^\dagger \phi_2)^2 + H.c.]
   + [ \lambda_6 (\phi_1^\dagger \phi_1)(\phi_1^\dagger \phi_2)
     + \lambda_7 (\phi_2^\dagger \phi_2)(\phi_1^\dagger \phi_2)
     + H.c.]
\nonumber
\\
&& - m_{11}^2 (\phi_1^\dagger \phi_1)
   - m_{22}^2 (\phi_2^\dagger \phi_2)
   - [ m_{12}^2 (\phi_1^\dagger \phi_2) + H.c. ],
\ee
where $\lambda_5, \lambda_6, \lambda_7$ and $m_{12}^2$ are
complex and others are real.
The discrete symmetry $\phi_1 \to -\phi_1$ or $\phi_2 \to -\phi_2$
leads to the absence of $m_{12}^2$, $\lambda_6$ and $\lambda_7$.
Here, we allow soft violation of the discrete symmetry by the dimension 2
terms $m_{12}^2 \ne 0$ to introduce the CP violation.

The minimization of the potential 
at $\langle \phi_1 \rangle = (0,v_1)^T/\sqrt{2}$
and $\langle \phi_2 \rangle = (0,v_2 e^{i \xi})^T/\sqrt{2}$
yields the relation:
\be
{\rm Im} ( m_{12}^2 e^{i \xi})
     = v_1 v_2 {\rm Im} ( \lambda_5 e^{2 i \xi} ),
\ee
where $v_1^2 + v_2^2 = v^2 = 2 m_W^2/g^2$ and
$\xi$ is the relative phase between $v_1$ and $v_2$.
By the rephasing invariance \cite{ginzburg},
we can choose $\xi=0$ which indicates no spontaneous CP violation
but the CP violation is entirely explicit.
Then the parameter ${\rm Im} ( m_{12}^2)$ can be replaced by
$ {\rm Im} ( \lambda_5)$ which is
the only CP violating parameter in this model.

The neutral states are defined by
$G^0 = \sqrt{2}
    ( {\rm Im}~ \phi_1^0 \cos \beta +  {\rm Im}~ \phi_2^0 \sin \beta )$,
$A^0 = \sqrt{2}
    ( -{\rm Im}~ \phi_1^0 \sin \beta +  {\rm Im}~ \phi_2^0 \cos \beta )$,
$\varphi_1 = \sqrt{2} {\rm Re}~ \phi_1^0$,
$\varphi_2 = \sqrt{2} {\rm Re}~ \phi_2^0$.
The $3 \times 3$ mass matrix of neutral Higgs bosons $(\varphi_1,\varphi_2,A)$ 
is constructed, of which $(1,3)$ and $(2,3)$ elements are given by
\be
{\cal M}^2_{13} = -\frac{1}{2} {\rm Im}~\lambda_5 v^2 \sin \beta,
~~~~~
{\cal M}^2_{23} = -\frac{1}{2} {\rm Im}~\lambda_5 v^2 \cos \beta,
\ee
where $\tan \beta = v_2/v_1$ .
Note that both ${\cal M}^2_{13}$ and ${\cal M}^2_{23}$
depend upon the parameter ${\rm Im}~\lambda_5$. 
These non-zero elements indicate mixing 
between the CP-even and CP-odd states
and imply the manifest CP violation.
We diagonalize the mass matrix by the orthogonal transformation
\be
{\cal M}^2_d = {\cal R} {\cal M}^2 {\cal R}^\dagger,
\ee
where the orthogonal matrix ${\cal R}$ is parametrized
by 3 Euler angles $\theta_a$,  $\theta_b$, $\theta_c$
\be
{\cal R} =
\left( \begin{array}{ccc}
-c_b s_a&c_a c_b&s_b \\
c_a c_c+s_a s_b s_c&s_a c_c-c_a s_b s_c&c_b s_c \\
-c_a s_c+ s_a s_b c_c&-s_a s_c-c_a s_b c_c&c_b c_c
           \end{array}   \right),
\ee
with $s_{a,b,c} = \sin \theta_{a,b,c}$
and $c_{a,b,c} = \cos \theta_{a,b,c}$.
Hereafter we set $\alpha \equiv \theta_a$ by convention.
Then the physical states for neutral Higgs bosons
$h_1, h_2, h_3$ are defined by
$(h_1,h_2,h_3)^T   = {\cal R} ( \varphi_1,\varphi_2,A)^T$.
The CP-odd state $A$ and CP-even states $\varphi_1, \varphi_2$
are no longer physical states
and it indicates a manifest CP violation in the neutral Higgs sector.

\section{The Scenarios}

The phenomenology of Higgs-gauge sector is governed by
the couplings of the Higgs bosons to gauge bosons.
The generalized $h_i ZZ$ couplings are given by
\be
h_1 ZZ &\sim& \sin (\beta-\alpha) \cos \theta_b,
\nonumber
\\
h_2 ZZ &\sim& \cos (\beta-\alpha) \cos \theta_c
              - \sin (\beta-\alpha) \sin \theta_b \sin \theta_c,
\nonumber
\\
h_3 ZZ &\sim& - \cos(\beta-\alpha) \sin \theta_c
              - \sin (\beta-\alpha) \sin \theta_b \cos \theta_c,
\ee
and the generalized $h_i h_j Z$ couplings given by
\be
Z h_1 h_3 &\sim& \cos (\beta-\alpha) \cos \theta_c
              - \sin (\beta-\alpha) \sin \theta_b \sin \theta_c,
\nonumber
\\
Z h_2 h_3 &\sim& -\sin (\beta-\alpha) \cos \theta_b,
\nonumber
\\
Z h_2 h_3 &\sim&  \cos(\beta-\alpha) \sin \theta_c
                + \sin (\beta-\alpha) \sin \theta_b \cos \theta_c,
\ee
which are normalized by the SM coupling $g m_Z/\cos \theta_W$.

If ${\rm Im}~ \lambda_5=0$,
the mass matrix is reduced to the CP conserving case,
where the imaginary parts and real parts of neutral scalar fields decouple.
This is corresponding to the case of $\theta_b = \theta_c = 0$
in the matrix ${\cal R}$.
In our general study, we consider other limiting cases
which could be of interest.
If we assume that $\theta_b \sim 0$ and $\theta_c \sim \pi/2$,
$h_2$ is decoupled and identified with the CP-odd Higgs boson $A$.
In the limit that $\theta_b \sim \theta_c \sim \pi/2$,
$h_1$ is decoupled to be $A$.
In both cases, the Higgs-gauge couplings $g_{h_i ZZ}$
and $g_{h_i h_j Z}$ go close to those of the CP conserving case.
Thus these limiting cases are similar to the CP conserving case
except that the CP-odd Higgs may be light.

More interesting scenario is obtained by taking the limit
$\sin \theta_c \to 0$.
In this case, the off-diagonal elements becomes
${\cal M}^2_{13} = s_a c_b s_b (m_3^2-m_2^2)$, and
${\cal M}^2_{23} = -c_a c_b s_b (m_3^2-m_2^2)$.
Considering the ratio ${\cal M}^2_{13}/{\cal M}^2_{23}$,
we obtain $\tan \beta = - \tan \alpha$.
Then the CP violating parameter ${\rm Im}~\lambda_5$
is directly related to $\theta_b$ and Higgs masses,
\be
{\rm Im}~\lambda_5 = \sin 2 \theta_b \frac{m_3^2-m_2^2}{v^2}.
\ee
If we additionally assume that $\sin \theta_b$ is close to 1,
the lightest Higgs decouples to be the CP-odd Higgs
and this limiting case may look like the CP conserving case
since the CP-odd Higgs decouples.
However we see that the ratio
$g_{h_2 ZZ}/g_{h_3 ZZ} = 1/ \tan (\beta-\alpha)$ in this case
while $g_{h ZZ}/g_{H ZZ}  = \tan (\beta-\alpha)$
in the CP conserving case.
It can be a signal to discriminate this scenario
from the CP conserving model in the gauge-Higgs sector
without manifest observation of the CP asymmetry.

     \begin{figure}[ht]
     \begin{center}
     \begin{tabular}{cc}
     \mbox{\epsfig{file=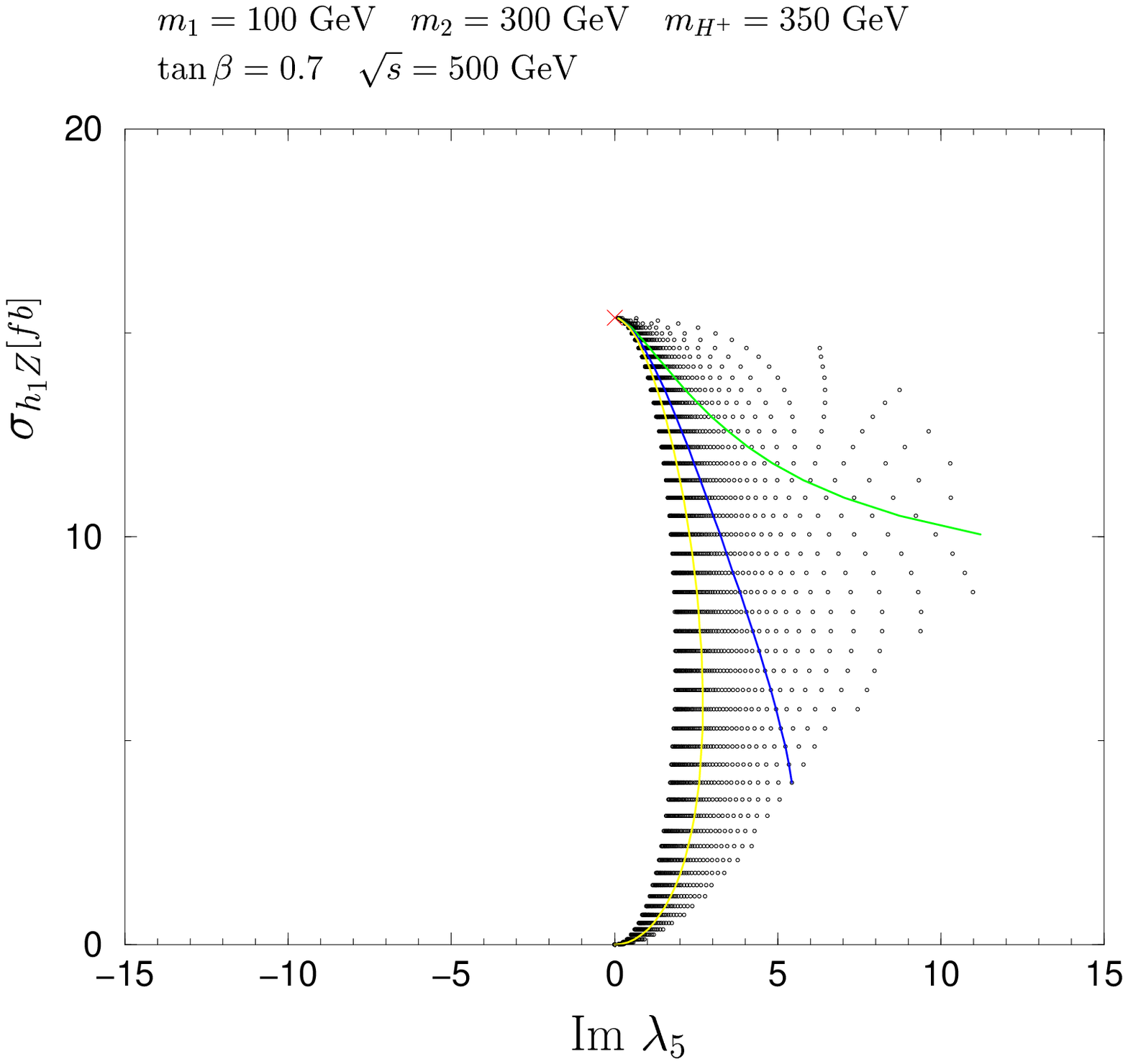,width=5.7cm}}&
     \mbox{\epsfig{file=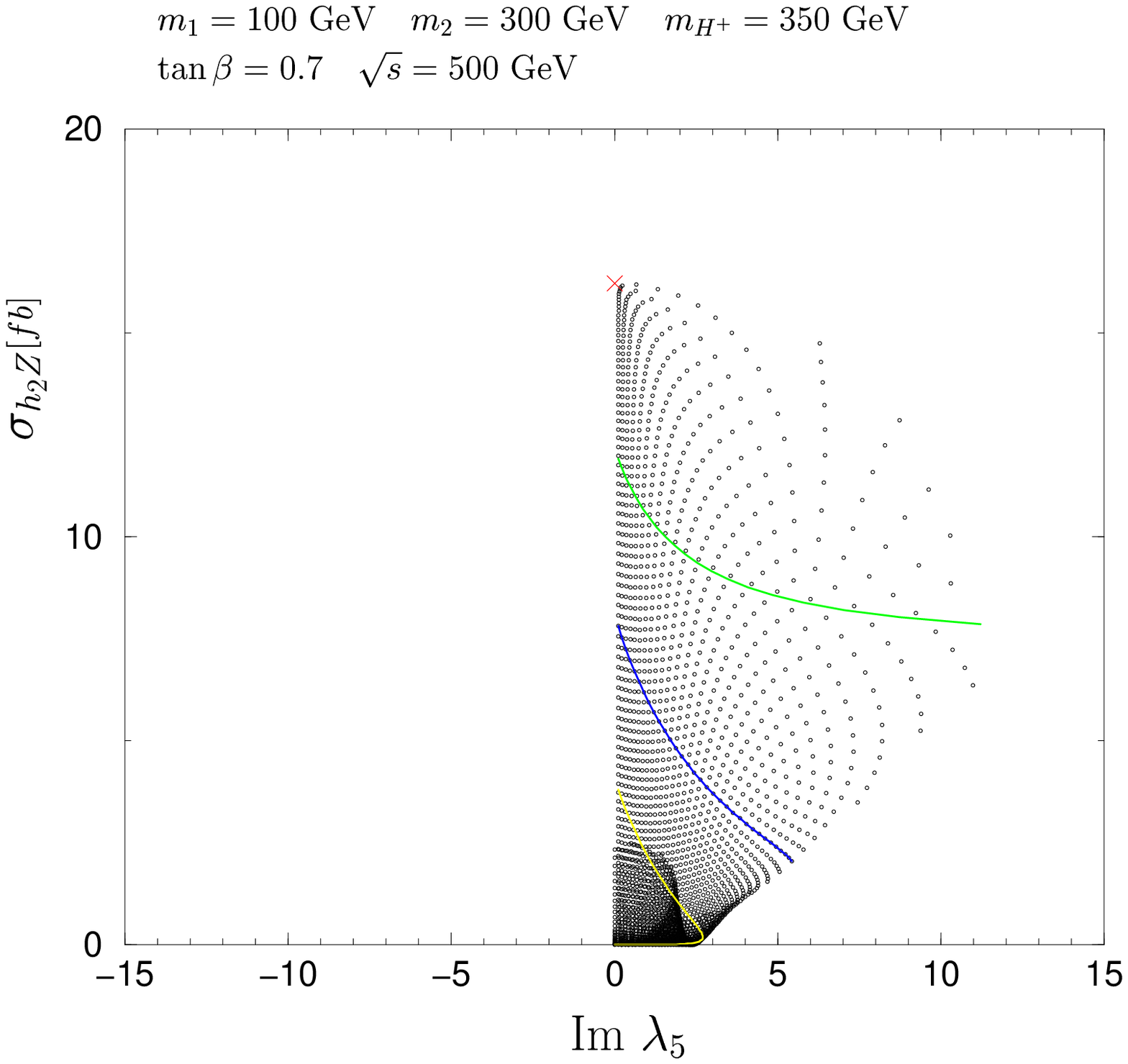 ,width=5.7cm}}
     \end{tabular}
     \end{center}
     \caption{Cross sectons for $e^- e^+ \to Z h_1$ 
                       and $e^- e^+ \to Z h_2$ processes}
     \label{fig:h1z}
     \end{figure}

\section{Neutral Higgs Boson Production}

The most promising channel for the neutral Higgs boson production
at the LC is the Higgsstrahlung process
$e^- e^+ \to Z h_i$.
For the numerical analysis,
we demand the following constraints on the model parameters \cite{maximal};
(1) the perturbativity on the quartic couplings, $\lambda_i/4 \pi <1$,
(2) the ordering of Higgs masses, $m_1 < m_2 <m_3$.
Figure 1 depicts the cross sections of $Z h_1$ and $Z h_2$ productions 
with respect to the CP violating parameter ${\rm Im}~\lambda_5$ 
when varying $\theta_b$ and $\theta_c$. 
The $\times$ marks denote the value of the CP conserving case 
and $\beta - \alpha$ is fixed to be $\pi/6$.
The green line denotes that $\theta_c=\pi/6$, the blue line $\theta_c=\pi/4$,
and the yellow line $\theta_c=\pi/3$.
We note that the cross section can be far away from that of the CP
conserving model even if ${\rm Im}~\lambda_5$ is close to 0.

In Fig. 2, we show the cross section for $e^- e^+ \to h_1 h_2$ process.
It can play a role of the supplementary process 
for the $e^- e^+ \to Z h_1$ channel
since the $Zh_1 h_2$ coupling becomes large when $ZZh_1$ coupling vanishes
\cite{gunion}.

     \begin{figure}[ht]
     \begin{center}
     \vspace*{.2cm}
     \epsfig{file=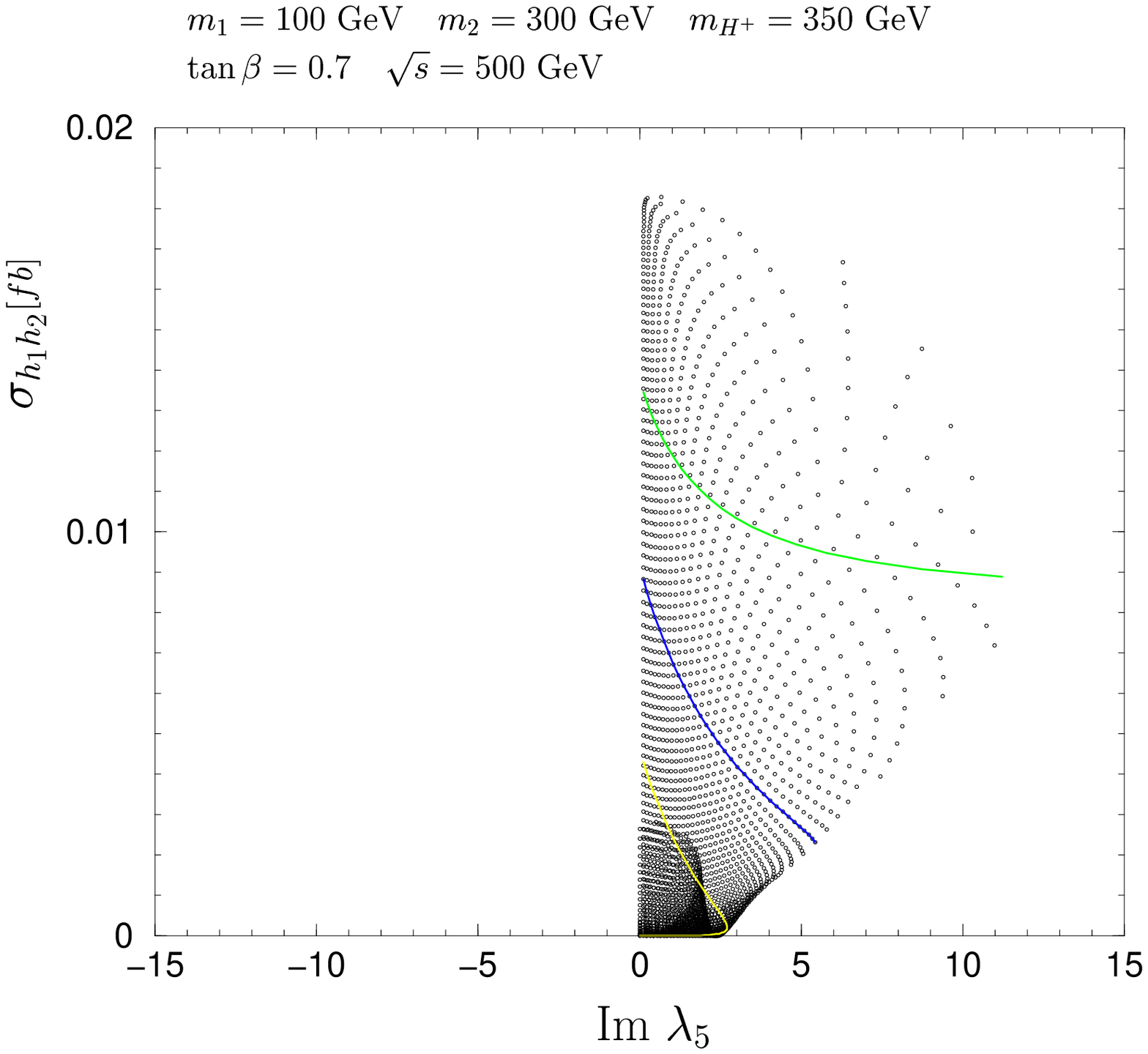,width=6.5cm}
     \end{center}
     \caption{Cross secton for $e^- e^+ \to h_1 h_2$ process}
     \label{fig:hh}
     \end{figure}

\section{Concluding Remarks}

The neutral Higgs boson production processes,
$e^- e^+ \to Z h_i$ and $e^- e^+ \to  h_i h_j$ at the LC
has been explored in the context of the two Higgs-doublet model
with CP violation.
We find that remarkably different phenomenology of the CP violating model 
is possible in the neutral Higgs boson production at the LC 
without direct CP violating signal.

\end{document}